\documentclass[aps,prl,reprint,superscriptaddress,showpacs,floatfix,longbibliography]{revtex4-1}

\usepackage{graphicx}
\usepackage{amssymb,amsmath}
\usepackage{dcolumn}
\usepackage{bm}
\usepackage[mathlines]{lineno}

\begin{document}
\title{Spontaneous antiferromagnetic order and strain effect on electronic properties of $\alpha$-graphyne}

\author{Baojuan Dong}
\affiliation{Shenyang National Laboratory
             for Materials Science,
             Institute of Metal Research,
             Chinese Academy of Sciences,
             University of Chinese Academy of Sciences,
             Shenyang 110016, China}
\author{Huaihong Guo}
\affiliation{College of Sciences, Liaoning Shihua University, Fushun, 113001, China}
\affiliation{Department of Physics, Tohoku University, Sendai 980-8578, Japan}
\author{Zhiyong Liu}
\affiliation{Shenyang National Laboratory
             for Materials Science,
             Institute of Metal Research,
             Chinese Academy of Sciences,
             University of Chinese Academy of Sciences,
             Shenyang 110016, China}
\author{Teng Yang}
\email
{yangteng@imr.ac.cn}%
\affiliation{Shenyang National Laboratory
             for Materials Science,
             Institute of Metal Research,
             Chinese Academy of Sciences,
             University of Chinese Academy of Sciences,
             Shenyang 110016, China}
\affiliation{Department of Physics, Tohoku University, Sendai 980-8578, Japan}
\author{Peng Tao}
\affiliation{Quanzhou Institute of Equipment Manufacturing, Haixi
Institutes, Chinese Academy of Sciences, Jinjiang, 362200, China}
\author{Sufang Tang}
\affiliation{Shenyang National Laboratory
             for Materials Science,
             Institute of Metal Research,
             Chinese Academy of Sciences,
             University of Chinese Academy of Sciences,
             Shenyang 110016, China}
\author{Riichiro Saito}
\affiliation{Department of Physics, Tohoku University, Sendai 980-8578, Japan}
\author{Zhidong Zhang}
\affiliation{Shenyang National Laboratory
             for Materials Science,
             Institute of Metal Research,
             Chinese Academy of Sciences,
             University of Chinese Academy of Sciences,
             Shenyang 110016, China}
\date{\today} 

\begin{abstract}
Using hybrid exchange-correlation functional in {\em ab initio} density functional theory calculations, we study magnetic
properties and strain effect on the electronic properties of $\alpha$-graphyne monolayer. We find that a spontaneous
antiferromagnetic (AF) ordering occurs with energy band gap ($\sim$ 0.5 eV) in the equilibrated $\alpha$-graphyne. Bi-axial
tensile strain enhances the stability of AF state as well as the staggered spin moment and value of the energy
gap. The antiferromagnetic semiconductor phase is quite robust against moderate carrier filling with threshold carrier
density up to 1.7$\times$10$^{14}$ electrons/cm$^2$ to destabilize the phase. The spontaneous AF ordering and strain
effect in $\alpha$-graphyne can be well described by the framework of the Hubbard model. Our study shows that it is
essential to consider the electronic correlation effect properly in $\alpha$-graphyne and may pave an avenue for exploring
magnetic ordering in other carbon allotropes with mixed hybridization of s and p orbitals.
\end{abstract}



\maketitle

%

\section*{Introduction}
Antiferromagnetic (AF) ordering in graphene induced by strain has recently attracted tremendous attention
\cite{Meng10,Wehling11,Lee12,Kotov12,Tang15,Roy14} and may supply a platform for both fundamental study
of the Coulomb interaction and possible applications on switchable magnetic devices. However, the
critical strain theoretically predicted\cite{Lee12} for a phase transition of graphene from semimetal to
AF semiconductor is around 8\%, which is practically not easy to achieve in experiment
and to demonstrate for potential applications. Graphyne as one type of graphene allotrope may serve as an
alternative to realize such transition at much lower strain.

\begin{figure}[t]
\includegraphics[width=1.0\columnwidth]{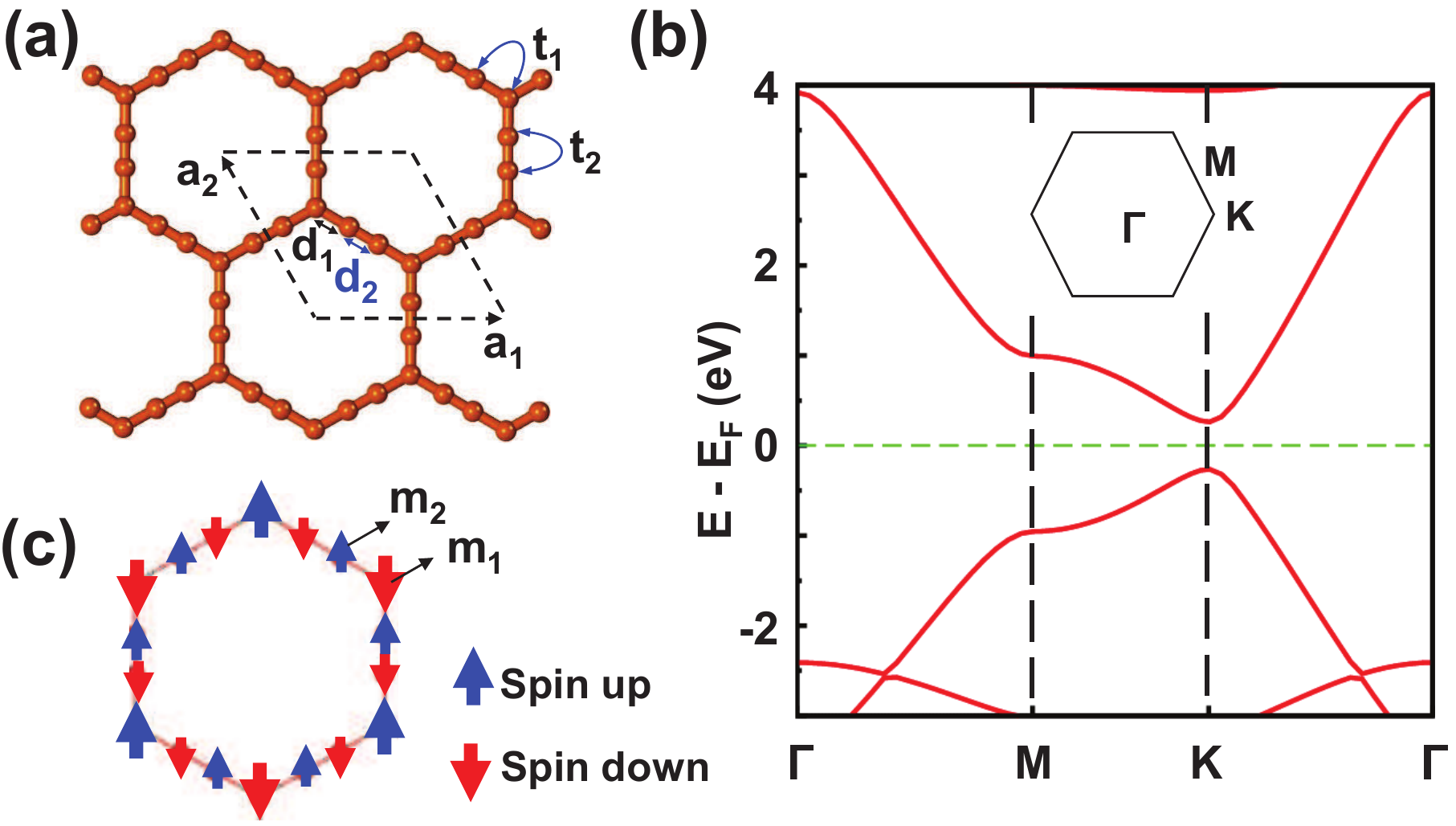}
\caption{(Color online) Spontaneous antiferromagnetic (AF) ordering of
$\alpha$-graphyne monolayer in the electronic ground-state. (a) Atomic
structure. (b) Energy band structure, bands for spin up and down are
degenerate. (c) Spin texture of $\alpha$-graphyne.
\label{Fig1} }
\end{figure}

Graphyne has been proposed several decades ago\cite{Baughman87}. Different from graphene with pure sp$^2$
hybridization, graphyne has both sp hybridization in the linear C-C bond and sp$^2$ hybridization at the hexagonal
corner (see Fig.~\ref{Fig1}(a)). Thus the variety in bonding states renders graphyne an appealing material
for studying the richness of electronic properties and tunability by strain. Recently, graphyne starts to
attract renewed and increasing attention, mainly due to some direction-dependent properties of massless
Dirac fermions and pseudospin state predicted in the graphyne systems\cite{Malko12,Malko12b,Kim12}. Worth
pointing out that most of the study on graphyne were purely
based on the single-particle picture without many-body effect taken into account. However, single-particle
picture has been proved insufficient, i.e., for the understanding of the Coulomb interaction in
graphene\cite{Sarma11} including the strain-induced AF state\cite{Tang15}, and especially in graphyne with
more localized sp states than the sp$^2$ states in graphene. We will show that a strong spin ordering occurs
in graphyne in contrast to graphene and that the Coulomb interaction in the structure with the sp C-C bond
can be controlled by much smaller strain than that for graphene, like polyacetylene\cite{Su79,Saito83},
which may lead to a smaller critical strain
or even zero strain for triggering spin-ordered state in graphyne.

To explore the phenomenon aforementioned, we studied electronic properties and strain effect in
$\alpha$-graphyne monolayer by first-principles calculations, which are further interpreted by the Hubbard model.
$\alpha$-graphyne has a similar hexagonal structure to graphene, but two extra carbon atoms are linearly inserted in
$\alpha$-graphyne between two nearest neighbor carbon atoms in graphene (see Fig.~\ref{Fig1}) . Here we selected $\alpha$-graphyne\cite{Baughman87}
in terms of the following considerations: (1) $\alpha$-graphyne
among all the proposed graphynes has the simplest structure with eight atoms per unit cell, but it still
remains the main character of the graphyne family that both sp$^2$ and sp hybridization coexist; (2) it has the
same point symmetry group (D$_{6h}$) as graphene, making a convenient analogy to graphene; (3) a model for
estimating the effective hopping integral $\tilde{t}$ in $\alpha$-graphyne has been proposed~\cite{Kim12},
and can facilitate an evaluation of the critical value of U/$\tilde{t}$ for a possible phase transition.
In this model, the so-called effective hopping integral $\tilde{t}$ between two carbon atoms at
the nearest hexagonal corner site in $\alpha$-graphyne can be described as -\textit{t$^2_1$/t$_2$} according
to Kim \textit{et al.}\cite{Kim12}, where \textit{t$_1$} and \textit{t$_2$} (\textit{t$_1$}, \textit{t$_2$} $>$ 0 as
shown in Fig.~\ref{Fig1}) are the hopping integral values for the single sp$^2$ bond and sp bond, respectively.
Since \textit{t$_1$} is due to sp$^2$ bond like C-C bond in graphene, \textit{t$_1$} can be approximately equal
to that in graphene. \textit{t$_2$} is due to sp bond which is shorter than sp$^2$ bond, and \textit{t$_2$} is larger than
\textit{t$_1$}\cite{Harrison}. Therefore, one expects that an effective $\tilde{t}$ in $\alpha$-graphyne is
smaller than the hopping integral value in graphene.
Such a small $\tilde{t}$ may be advantageous for a semimetal-antiferromagnetic semiconductor transition in
$\alpha$-graphyne occurring under a reduced strain, since the kinetic exchange~\cite{Anderson} is generally
expressed by kinetic exchange interaction, $\tilde{t}^2$/U, in which U is the on-site Coulomb interactions.

\begin{figure*}[t]
\includegraphics[width=1.8\columnwidth]{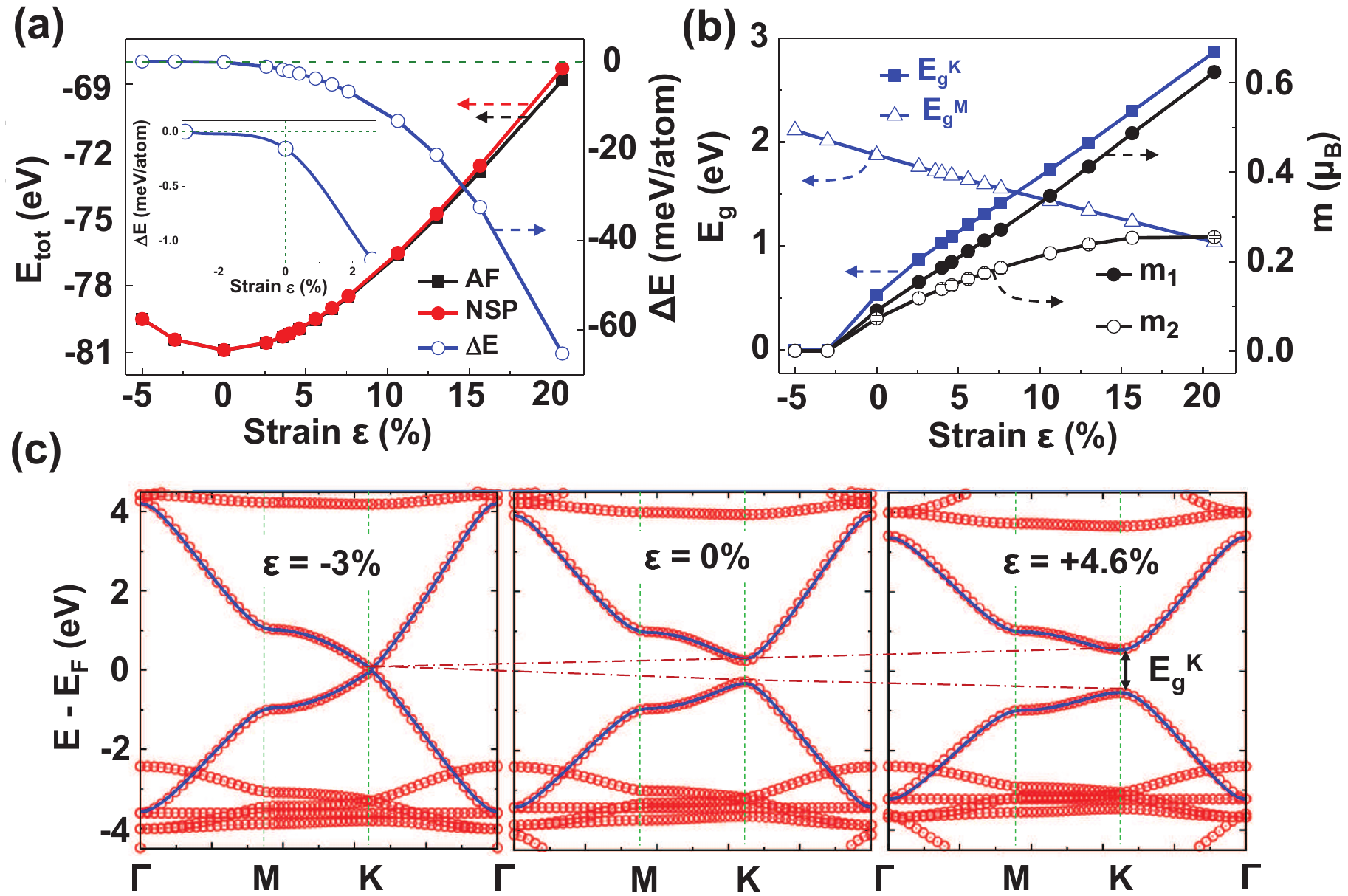}
\caption{(Color online) Strain effect on $\alpha$-graphyne. (a) The stability of antiferromagnetic ordering state (AF)
with respect to non-polarized state (NSP) as a function of strain $\varepsilon$. The energy for the AF and NSP states are
showed in black solid square and blue solid circle, respectively. And the energy difference $\Delta$E (= E$_{tot}$(AF)
- E$_{tot}$(NSP)) is showed in blue empty circle. (b) Energy band gap E$^M_g$ and E$^K_g$, and staggered spin moment
\textit{m}$_1$ and \textit{m}$_2$ at two inequivalent atomic sites as a function of $\varepsilon$. (c) Band structures at
three strains $\varepsilon$ = $-$3\%, 0\% and $+$4.6\%. The empty circles represent the band structure from DFT calculations,
and the blue solid lines are from the Hubbard model calculations which is to fit the DFT results. The brown dashed lines
are used to highlight the band gap E$_g^K$ evolving with strain.
\label{Fig2}}
\end{figure*}
To evaluate the electron-electron interaction in $\alpha$-graphyne, we used hybrid exchange-correlation functional~\cite{HSE03}
in {\em ab initio} density functional theory calculations. We found that the critical strain ($\sim$ $-$3\%, compressive)
for a semimetal-AF semiconductor transition in $\alpha$-graphyne is much reduced than that ($\sim$ $+$8\%, tensile)
of graphene. And more importantly, spontaneous AF spin ordering of semiconducting electronic ground state appears even
at zero strain. Bi-axial tensile strain can enhance the stability of the AF state as well as the staggered spin moment
and energy band gap at the zone-corner K point in the hexagonal Brillouin zone.  The antiferromagnetic semiconductor
phase is destabilized on carrier filling with a threshold carrier doping density up to 1.7$\times$10$^{14}$
electrons/cm$^2$. A much smaller effective hopping integral in $\alpha$-graphyne than in graphene is
responsible for the spontaneous AF ordering as is understood by the Hubbard model. The strain-enhanced stability
of AF ground state in $\alpha$-graphyne may be observed experimentally even at finite temperature.

\section*{Computational Techniques}

In order to obtain the ground state and strain-induced properties in $\alpha$-graphyne, we used {\em ab initio}
density functional theory as implemented in the \textsc{VASP} code\cite{VASP}. We use a periodic boundary condition
with monolayer structures represented by a periodic array of slabs separated by a vacuum region (${\agt}$~19.5~{\AA}).
We use the projector augmented wave (PAW) pseudopotentials~\cite{PAWPseudo} and the Perdew-Burke-Ernzerhof (PBE)~\cite{PBE}
exchange-correlation functional. The Brillouin zone of the primitive unit cell of the 2D structures is sampled by
$7{\times}7{\times}1$~$k$-points~\cite{Monkhorst-Pack76}. We adopt $500$~eV as the electronic kinetic energy cutoff
for the plane-wave basis and $10^{-6}$~eV for a total energy difference between subsequent self-consistency iterations
as the criterion for reaching self-consistency. All geometries are optimized using the conjugate gradient method~\cite{CGmethod},
until none of the residual Hellmann-Feynman forces exceeds $2{\times}10^{-2}$~eV/{\AA}. Uniform strain $\varepsilon$
used in this study is defined as \textit{(a-a$_0$)/a$_0$}, in which \textit{a$_0$} and \textit{a} are, respectively,
the lattice constants without and with a strain.

To remedy the self-interaction error of the GGA calculations in carbon allotropes, we further performed the hybrid-functional
(HSE06) calculations~\cite{HSE03} to determine more accurately the ground state and strain effect. In the calculations,
the PBE exchange energy and Hartree-Fock exchange energy were hybridized, along with the full PBE correlation energy.


The Hubbard model with Hartree-Fock mean-field approximation is also adopted here to get an essence of the emergent spontaneous AF order in $\alpha$-graphyne system.

\begin{eqnarray}
  \hat{H}_{HF} &=& - \sum_{i j \sigma} \sum_{<p, q>} t^{i,j}_{p,q} \hat{c}^+_{i p \sigma} \hat{c}_{j q \sigma} \nonumber \\
  &+& U \sum_{i p \sigma} \left [  n_{i p \overline{\sigma}} \hat{c}^+_{i p \sigma} \hat{c}_{i p \sigma}  -\frac{1}{2} n_{i p \overline{\sigma}} n_{i p \sigma} \right ]
\label{eq1}
\end{eqnarray}
in which \emph{t}$^{i,j}_{p,q}$ represents hopping integral between the p$^{th}$ atom of the i$^{th}$ unit cell and the q$^{th}$ atom of the
j$^{th}$ unit cell, U the Coulomb repulsion, i,j the index of lattice vectors, $\sigma$ spin index, p, q the index of atoms in the unit cell.
Since the system is homogeneous, we can assume that number of particle at each atomic site has translational symmetry, n$_{i p \sigma}$ $\equiv$
n$_{p \sigma}$. Then, n$_{p \sigma}$ is expressed by
\begin{eqnarray}
    n_{p \sigma} &=& \left< \hat{c}^+_{i p \sigma} \hat{c}_{i p \sigma} \right> \nonumber \\
    &=& \frac{1}{N} \sum_i \left< \hat{c}^+_{i p \sigma} \hat{c}_{i p \sigma} \right> \nonumber \\
    &=& \frac{1}{N} \sum_k \left< \hat{c}^+_{k p \sigma} \hat{c}_{k p \sigma} \right> ,
\label{eq2}
\end{eqnarray}
here $N$ is the number of k points in the 1$^{st}$ Brillouine zone, and $\hat{c}^+_{k p \sigma}$ ($\hat{c}_{k p \sigma}$) is the Fourier transform of
$\hat{c}^+_{i p \sigma}$ ($\hat{c}_{i p \sigma}$).

\section*{results and discussion}
Figure~\ref{Fig1} shows the calculated results of $\alpha$-graphyne at zero strain ($\varepsilon$ = 0).
The optimized structure in Fig.~\ref{Fig1}(a) shows two different bonding
length defined as \textit{d$_1$} = 1.39 \AA\ and \textit{d$_2$} = 1.22 \AA\ for the
two distinct hybridizations sp$^2$ and sp,
respectively. In Fig.~\ref{Fig1}(b) we show the calculated electronic energy band.
As shown in Fig.~\ref{Fig1}(b), $\alpha$-graphyne is a semiconductor with
a direct energy band gap of around 0.5 eV occurring at the zone corner K point of
Brillouin zone. AF ordering appears with the degenerate electronic energy bands
for spin up and down over the Brillouin zone. The inversion symmetry of the unit
cell is broken by such an antiferromagnetic (AF) ordering, as shown in Fig.~\ref{Fig1}(c).
The magnetic moment \textit{m} for the corner carbon (\textit{$m_1$}) is slightly larger
than that of the edge
carbons (\textit{$m_2$}), as shown in Fig.~\ref{Fig2}(b). The calculated results show that
we have obtained both AF and semiconducting state in the $\alpha$-graphyne for $\varepsilon$ = 0
when taking the electronic correlation into account. While for graphene at zero strain there is no
magnetic order, except for the boundary atoms at the zigzag edges~\cite{Son06,Son06b,G-M14}.

\begin{figure}[t]
\includegraphics[width=1.0\columnwidth]{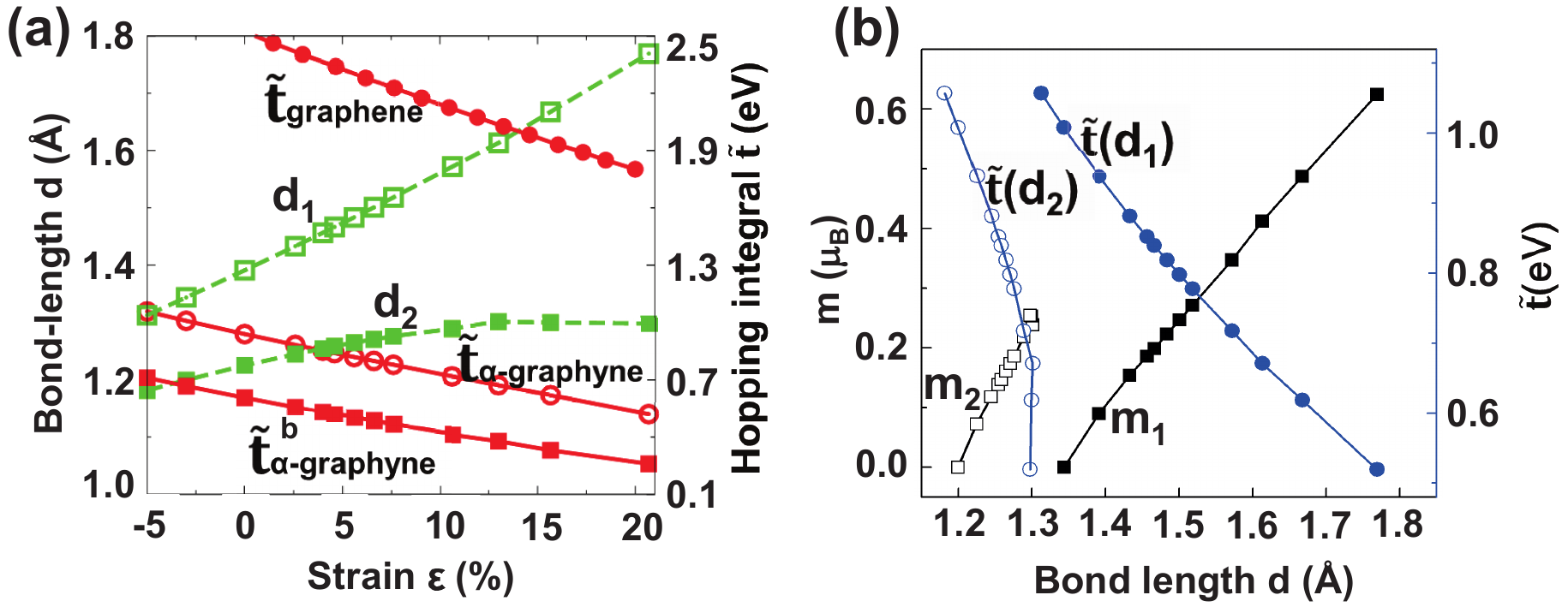}
\caption{(Color online) (a) Strain-dependent bond lengths (d$_1$ and d$_2$ defined in Fig.~\ref{Fig1}(a))
and effective hopping integral ($\tilde{t}$) in $\alpha$-graphyne. Data of graphene in filled circles in
(a) is from Lee et al.\cite{Lee12} and used for a compare, $\tilde{t}_{\alpha-graphyne}$ and
$\tilde{t}^b_{\alpha-graphyne}$ are derived from two ways (refer to the main text for details).
$\tilde{t}_{\alpha-graphyne}$ from half of E$_g^M$ is used for more discussions.
(b) $\tilde{t}$ and staggered magnetic moment as a function of bond lengths.
\label{Fig3}}
\end{figure}

To understand the spontaneous AF ordering, we applied uniform strain to $\alpha$-graphyne
as a perturbation and calculated the strain effect on the electronic properties in $\alpha$-graphyne.
In Fig.~\ref{Fig2}(a) we show the calculated total energies E$_{\textrm{tot}}$ of both non-spin-polarized (NSP)
and AF states and also the energy difference $\Delta$E (= E$_{\textrm{tot}}$(AF) - E$_{\textrm{tot}}$(NSP))
as a function of strain $\varepsilon$. Total energy minimum at zero strain indicates that
the atomic structure is well optimized. From the inset of Fig.~\ref{Fig2}(a), we can
see that $\Delta$E at zero strain is not zero but around 0.2 meV/atom, indicating that NSP state
is less stable than the AF one. The energy difference becomes more pronounced with the increasing
$\varepsilon$ and obviously tensile strain ($\varepsilon >$ 0) can enhance the relative
stability of AF to the NSP state, which may be understood from the increased Peierls distortion
between sp and sp$^2$ bond lengths due to tensile strain, as seen from \textit{d$_1$} and \textit{d$_2$}
changing with $\varepsilon$ in Fig.~\ref{Fig3}(a). In contrast to the effect due to a tensile strain,
a compressive strain ($\varepsilon <$ 0) is found to reduce $\Delta$E till zero at $\varepsilon\sim$ $-$3.0\%.

Strain changes not only the relative stability of spin-polarized state, but also the
magnitude of the band gap E$_g$ and spin moment \textit{m}$_1$ and \textit{m}$_2$ in the
semiconducting AF state. Figure~\ref{Fig2}(b) shows the strain dependence of E$_g$ (blue solid
square and open triangle) at the K and M points, respectively, and \textit{m}$_1$ and
\textit{m}$_2$ (black solid and open circles). The E$^K_g$ at the K point is a fundamental
band gap of 0.5 eV at $\varepsilon$ = 0. A compressive strain at $\varepsilon\sim$ $-$3\% closes the
band gap, while a tensile strain increases E$^K_g$ almost linearly. This trend is also
shown in the strain-dependent electronic band structures in Fig.~\ref{Fig2}(c). Two dashed
lines are used to highlight the band gap E$^K_g$ changing with strain. On the other hand,
E$^M_g$ decreases with increasing tensile strain, which may have something to do with the
electronic hopping which decreases with bond elongation due to tensile strain, as discussed
more in Fig.~\ref{Fig3}. The spin magnetization has a similar trend as E$^K_g$. At zero strain,
\textit{m}$_1$ and \textit{m}$_2$ both have
finite value ($\sim$ 0.1 $\mu_B$) and go to zero at $\varepsilon\sim$ $-$3.0\% where E$^K_g$
drops to zero. Both \textit{m}$_1$ and \textit{m}$_2$ also increase with tensile strain.
\textit{m}$_2$ increases less rapidly than \textit{m}$_1$ with increasing $\varepsilon$ and
saturates around 0.25 $\mu_B$
at $\varepsilon$ $\sim$ $+$20\%. In contrast, \textit{m}$_1$ seems to increase linearly with strain
and has no saturation. The strain dependence of spin magnetization \textit{m}$_1$ (\textit{m}$_2$)
is the same as that of bond length \textit{d$_1$} (\textit{d$_2$}) of sp$^2$ (sp) bond, as shown in green in
Fig.~\ref{Fig3}(a) and also as indicated by the linear dependence of magnetization \textit{m}$_1$
(\textit{m}$_2$) on bond length \textit{d$_1$} (\textit{d$_2$}) in black dots in Fig.~\ref{Fig3}(b), suggesting
contribution of strain-enhanced electron localization to the spin magnetization.

The localization of electrons with increasing tensile strain can be seen from the hopping integral values, too.
Hopping integrals are parameters which monotonically decrease with increasing bond-length~\cite{Harrison}.
Figure~\ref{Fig3}(a) shows the bond-lengths \textit{d$_1$} and \textit{d$_2$} as a
function of strain. Here \textit{d$_1$} and \textit{d$_2$} are defined in Fig.~\ref{Fig1}(a) as the bond-length of sp$^2$
and sp bonds, respectively. \textit{d$_1$} increases with strain more abruptly than \textit{d$_2$}, indicating the
hopping integral t$_1$ of the sp$^2$ bond drops more substantially than t$_2$ of the sp bond.
t$_1$ should be similar to the value in graphene due to a similar $\pi_z$ electron hopping on top
of sp$^2$ hybridization. t$_2$ has contributions from both $\pi_x$ and $\pi_y$ hopping channels
on top of sp hybridization and is expected to be larger than t$_1$. The effective hopping integral
$\tilde{t}$ from one hexagonal
corner site to another is therefore expected to smaller than t$_1$ and decreases with increasing tensile
strain faster than t$_1$. The calculated hopping integral confirms this trend, as shown
in Fig.~\ref{Fig3}(a). The value of effective hopping integral $\tilde{t}$ is extracted from two
ways, one of which ($\tilde{t}_{\alpha-graphyne}$ in Fig.~\ref{Fig3}(a)) is from the band gap
E$_g^M$ (E$_g^M$ = 2$\tilde{t}$) at the M point (the zone edge center point of hexagonal Brillouin
zone), the other ($\tilde{t}^b_{\alpha-graphyne}$ in Fig.~\ref{Fig3}(a)) based on $\tilde{t}$
($\sim$ -t$^2_1$/t$_2$)~\cite{Kim12} and t$_i$ $\sim$ -0.63$\frac{\hbar^2}{m d_i^2}$~\cite{Harrison}.
From Fig.~\ref{Fig3}(a), $\tilde{t}$s of $\alpha$-graphyne extracted from two ways are slightly
different, but are both much less than that of graphene.

\begin{figure}[b]
\includegraphics[width=0.80\columnwidth]{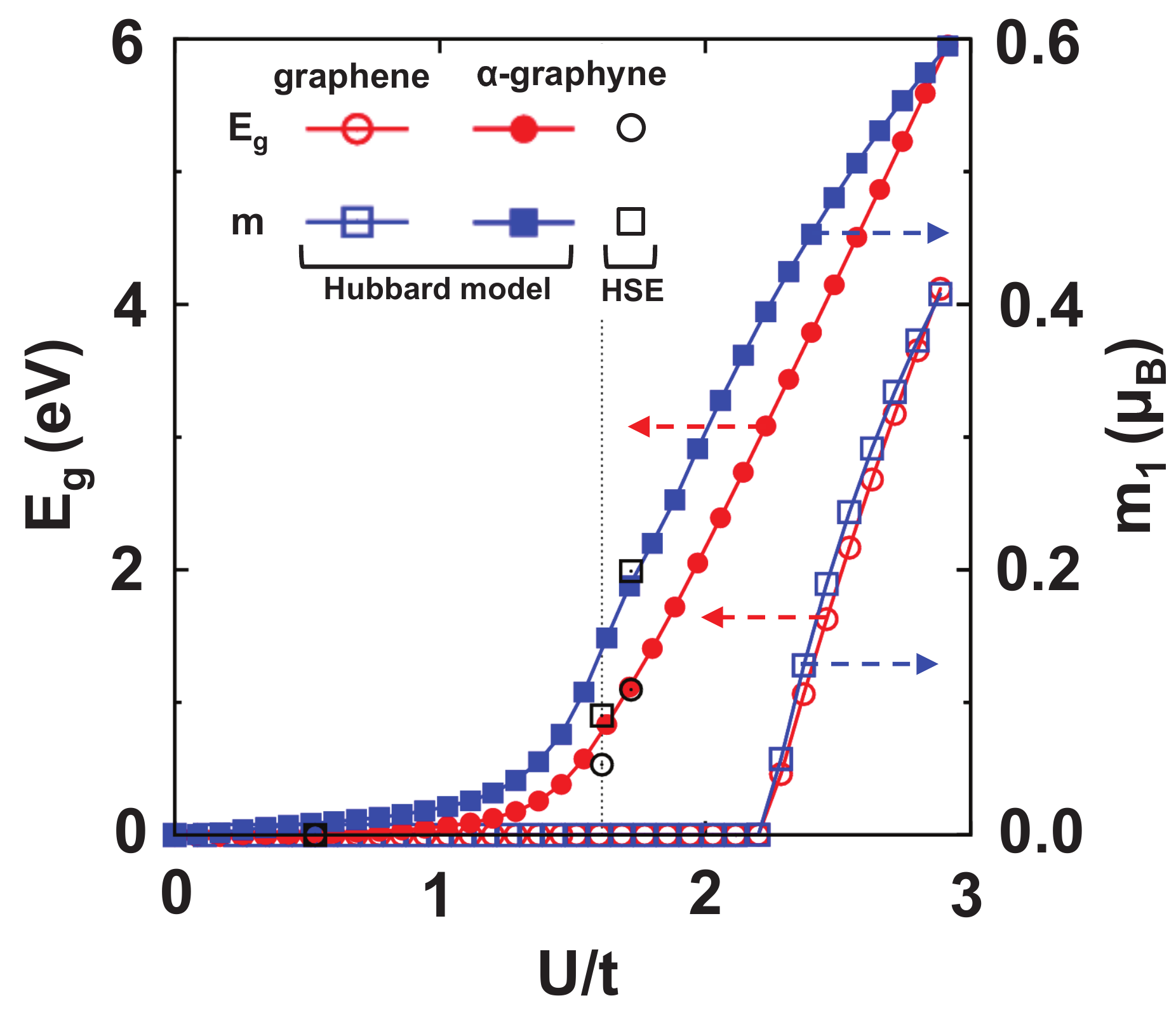}
\caption{(Color online) Energy band gap and magnetization m$_1$ as a function of U/t in $\alpha$-graphyne
from Hubbard Model calculations. Graphene result from Hubbard Model calculations (blue open square
and red open circles) and $\alpha$-graphyne from hybrid functional (HSE) calculations (black open square
and circles) are also given for a comparison. A vertical dotted line at U/t = 1.61 is to show the
HSE result of $\alpha$-graphyne at $\varepsilon$ = 0.
\label{Fig4}}
\end{figure}

In order to understand the phenomena generally, we consider the Hubbard model (Eq.~\ref{eq1}). The Hubbard
model has been widely used to understand the strain-induced semimetal-antiferromagnetic semiconductor
transition in graphene~\cite{Kim12,Tang15,Roy14,Wehling11}. The electron-electron on-site Coulomb interaction U
competes with electronic hopping integral t. Monte-Carlo calculations \cite{Meng10} based
on the Hubbard model have predicted a critical value of U/t in graphene ($\sim$4.3), over which energy gain due to
electronic hopping can not counteract energy cost due to the Coulomb repulsion and transition occurs. Within the
Hartree-Fock mean-field approximation, we solved the Hubbard model of $\alpha$-graphyne and also of graphene
for comparison. Figure~\ref{Fig4} shows band gap E$_g$ and magnetization \textit{m$_1$} as a function of U/t
for both $\alpha$-graphyne and graphene. We obtained the same critical value (U/t)$_c$ in graphene as in literature
using the same method~\cite{Sorella92,Martelo96}. It is reasonable that (U/t)$_c$ ($\sim$2.2) for graphene from Hatree-Fock
mean field lies below the value ((U/t)$_c\sim$4.3) from Monte-Carlo calculations\cite{Kim12,Tang15,Wehling11}.
For comparison, $\alpha$-graphyne shows a similar trend of both band gap and magnetization with U/t as graphene,
but $\alpha$-graphyne has a much reduced and ill-defined critical value ( (U/t)$_c\sim$1.5 from a linear
extrapolation of the band gap at large U/t), suggesting an onset of transition to the AF state at much smaller strain cost than
in graphene. By fitting the band structures of Hubbard model to the HSE band structures near the Fermi level in
Fig.~\ref{Fig2}(c), we obtained the U/t values of 0.53, 1.61, 1.72 at a strain of $-$3\%, 0\%, $+$4.6\%, respectively
and showed the three data points in black open square and circles in Fig.~\ref{Fig4}. In graphene with high hopping integral,
a sizable tensile strain ($\sim$ $+$8\%) is predicted to be necessary for U/t to approach the critical value
($\sim$2.2)~\cite{Kim12,Tang15,Wehling11}, whereas compared with graphene, in $\alpha$-graphyne electronic hopping integral
is much smaller, getting U/$\tilde{t}$ in $\alpha$-graphyne at zero strain (A vertical dotted line in Fig.~\ref{Fig4})
already over the critical value and therefore the spontaneous antiferromagnetic semiconducting electronic ground state
occurs. Further, when U/$\tilde{t}$ increases with tensile strain, it stabilizes the AF state. The strain-enhanced
stability of the AF ground state in $\alpha$-graphyne is advantageous for experimentalists to test this emergent phenomenon
at finite temperature.

\begin{figure}[t]
\includegraphics[width=1.0\columnwidth]{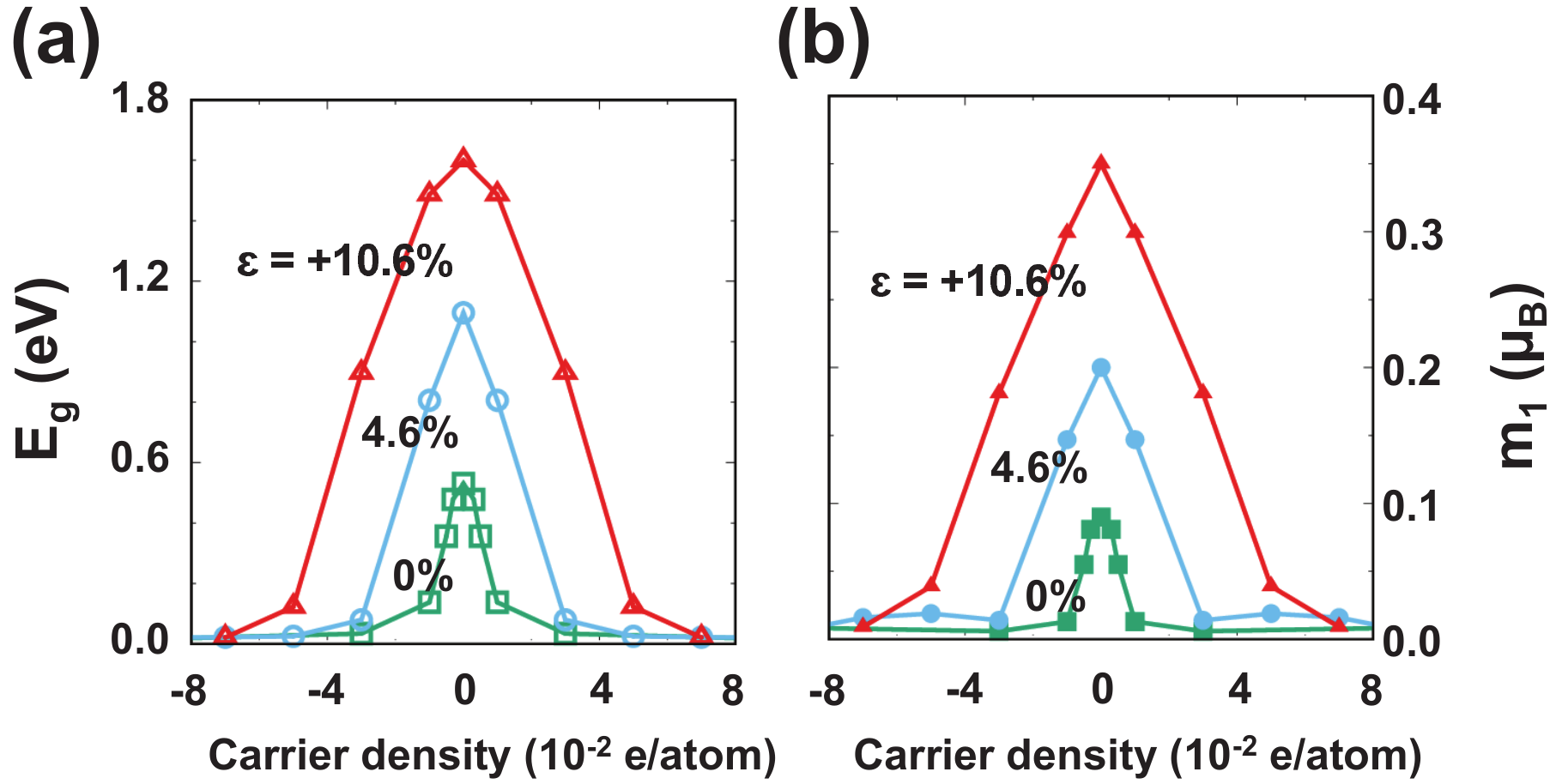}
\caption{(Color online) Carrier doping effect in $\alpha$-graphyne. (a) The band gap and (b) spin
moment \textit{m}$_1$ as a function of carrier density for tensile strain $\varepsilon$ from 0\% to 10.6\%.
\label{Fig5}}
\end{figure}

The strain-enhanced stability of the spontaneous AF insulator phase is also reflected in a competition
between tensile strain and carrier doping. Carrier by either filling conduction band or emptying valence band tends
to close the band gap and reduce spin magnetic moment, as shown in Fig.~\ref{Fig5}(a,b), since the energy gain by
opening the energy gap is not expected. At zero strain, both band gap E$_g$ and spin moment \textit{m$_1$} decrease with
increasing (electron or hole) carrier density and approach zero at 0.03 electrons/atom (or a threshold carrier
density at equivalently 1.7$\times$10$^{14}$ electrons/cm$^2$). This threshold carrier density is quite large compared
with the value of graphene~\cite{Lee12}, indicating a good stability of antiferromagnetic state against the doping effect.
The threshold value for E$_g$ and \textit{m$_1$} increases with tensile strain as shown in Fig.~\ref{Fig5}. Moreover,
the way of E$_g$ and \textit{m$_1$} approaching zero is different from that of graphene.
Lee \emph{et al.}~\cite{Lee12} found a bell-shaped decrease upon deviation from the band gap maximum in graphene, whereas in $\alpha$-graphyne
we found a hat-shaped decrease of both E$_g$ and \textit{m$_1$} from their maxima, in a much slower pace to get to zero.
This hat-shaped behavior of both E$_g$ and \textit{m$_1$} with carrier density in Fig.~\ref{Fig5} is the same as that
with U/t in Fig.~\ref{Fig4}, suggesting that the phase transition from AF semiconductor to semi-metal in $\alpha$-graphyne
is second-order.

In summary, we show a spontaneous AF semiconducting state in $\alpha$-graphyne by using hybrid functional
calculations. Strain can increase the stability, band gap size at the K point and magnetic spin moments at
each atomic site. The electronic hopping integral $\tilde{t}$ in $\alpha$-graphyne is found to be much smaller
than that in graphene, which is essential to the understanding of such unusual spontaneous AF spin ordering.
$\tilde{t}$ goes down with tensile strain, making AF state more robust and easier to test in experiment at
non-zero temperature. The spontaneous AF insulator is quite robust against carrier doping effect, generating
a hat-shaped decrease of both E$_g$ and spin moment in $\alpha$-graphyne other than a bell-shaped decrease
found in graphene. This study can be investigated by experiments in which strain and doping can be controlled
systematically.

\begin{acknowledgments}
This project is supported by the National Key R\&D Program of China (No.2017YFA0206301) and the Major
Program of Aerospace Advanced Manufacturing Technology Research Foundation NSFC and CASC, China (No. U1537204).
R.S. acknowledges JSPS KAKENH Grant number JP25107005 and JP15K21722. H.G. acknowledges NSFC Grant No. 51702146,
College Students' innovation and entrepreneurship projects (No. 201710148000072) and Liaoning Province Doctor
Startup Fund (No. 201601325). B.J.D. and T.Y. acknowledge China Scholarship Council for financial support.
\end{acknowledgments}

%


\end{document}